\documentclass[a4paper,numcites,final]{aipproc}
\setcitestyle{numbers}
\layoutstyle{8x11double}

\usepackage{amsmath}
\DeclareMathOperator{\Det}{Det}	
\newcommand*{\be}{\begin{equation}}
\newcommand*{\ee}{\end{equation}}
\newcommand*{\vp}{{\vec{p}}}
\newcommand*{\vq}{{\vec{q}}}
\newcommand*{\vk}{{\vec{k}}}
\newcommand*{\calD}{\ensuremath{\mathcal{D}}}
\newcommand*{\Eqref}[1]{Eq.\ \eqref{#1}}
\newcommand*{\vev}[1]{\left< #1 \right>}
\providecommand*{\coloneq}{\mathrel{\mathop:}=}
\providecommand*{\eqcolon}{=\mathrel{\mathop:}}
\newcommand*{\abs}[1]{\ensuremath{\lvert#1\rvert}}
\newcommand*{\e}{\ensuremath{\mathrm{e}}}
\renewcommand*{\i}{\ensuremath{\mathrm{i}}}

\begin{document}

\title{Variational Approach to Yang--Mills Theory with non-Gaussian Wave Functionals}

\classification{11.10.Ef, 12.38.Aw, 12.38.Lg}
\keywords{Coulomb gauge, variational techniques}

\author{Davide R.\ Campagnari}{address={Institut f\"ur theoretische Physik, Universit\"at T\"ubingen,
Auf der Morgenstelle 14, 72076 T\"ubingen, Germany}}

\author{Hugo Reinhardt}{address={Institut f\"ur theoretische Physik, Universit\"at T\"ubingen,
Auf der Morgenstelle 14, 72076 T\"ubingen, Germany}}

\begin{abstract}
A general method for treating non-Gaussian wave functionals in quantum field theory is
presented and applied to the Hamiltonian approach to Yang-Mills theory in Coulomb gauge
in order to include a three-gluon kernel in the exponential of the vacuum wave functional.
The three-gluon vertex is calculated using the propagators found in the variational
approach with a Gaussian trial wave functional as input.
\end{abstract}

\maketitle


\section{Introduction}

Over the last few years, there have been substantial efforts devoted to a variational
solution of the Yang--Mills Schr\"odinger equation in Coulomb gauge \citep{SzcSwa01,FeuRei04,EppReiSch07}.
In this approach,
using Gaussian type wave functionals, minimization of the energy density results in the
so-called gap equation for the inverse equal-time gluon propagator. This equation
has been solved analytically in the ultraviolet \citep{FeuRei04} and in the infrared \citep{SchLedRei06},
and numerically in the full momentum regime \citep{FeuRei04,EppReiSch07}.
One finds an inverse gluon propagator which in the UV behaves like the
photon energy but diverges in the IR, signalling confinement. The obtained
propagator also compares favourably with the available lattice data. There are, however,
deviations in the mid-momentum regime (and minor ones in the UV) which can be attributed
to the missing gluon loop, which escapes the Gaussian wave functionals. These deviations
are presumably irrelevant for the confinement properties, which are dominated by the ghost
loop (which is fully included under the Gaussian ansatz), but are believed to be important
for a correct description of spontaneous breaking of chiral symmetry \citep{YamSug10}. 

In this talk, we present a generalization of the variational approach to the
Hamiltonian formulation of Yang--Mills theory \cite{FeuRei04} to non-Gaussian wave functionals. The
expectation value of the Hamilton operator can be expressed in terms of the variational
kernels occurring in the ansatz through Dyson--Schwinger equations (DSEs). The three-gluon
vertex and the effects of the gluon loop on the gluon propagator are investigated.


\section{Non-Gaussian wave functionals}

\begin{figure}
\includegraphics[width=\linewidth]{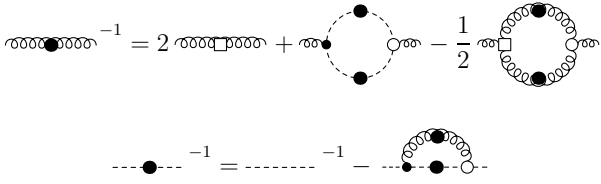}
\caption{\label{fig:dses}Dyson--Schwinger equation for the gluon (top) and ghost propagator
(bottom). Here and in the following, small filled dots represent propagators, small empty
dots vertex functions, and empty boxes the variational kernels.}
\end{figure}

In the Hamiltonian approach to Yang--Mills theory in Coulomb gauge, the vacuum expectation
value (VEV) of an operator depending on the transverse gauge field is given by
\be\label{mad1}
\vev{K[A]} = \int \calD A \: J_A \: \abs{\psi[A]}^2 \: K[A] ,
\ee
where $J_A=\Det(G_A)$ is the Faddeev--Popov determinant, $G_A=(-\partial\Hat{D})^{-1}$
is the inverse Faddeev--Popov operator, and $\psi[A]$ is the vacuum wave functional. In
\Eqref{mad1}, the functional integration runs over transverse field configurations
$\partial_i A_i^a=0$ and is restricted to the first Gribov region. Writing the vacuum
functional as
\be\label{mad2}
\abs{\psi[A]}^2 \eqcolon \exp\{-S[A]\} ,
\ee
one can derive DSEs from the functional identity
\be\label{mad3}
0 = \int \calD A \: \frac{\delta}{\delta A} \bigl\{ J_A \: \e^{-S[A]} \: K[A] \bigr\} .
\ee
In this talk, we consider a functional of the form
\be\label{mad4}
S[A] = \omega A^2 + \frac{1}{3!} \: \gamma_3 \, A^3 ,
\ee
where $\omega$ and $\gamma_3$ are variational functions.
A vacuum functional containing also a quartic term is discussed in Ref.\ \citep{CamRei10}.
With the explicit form \Eqref{mad4} for the vacuum functional, the DSEs are the usual
DSEs of Landau gauge Yang--Mills theory, however, in $D=3$ dimensions and with the bare
vertices of the usual Yang--Mills action replaced by the variational kernels. For the
gluon $\vev{AA}\eqcolon1/(2\Omega)$ and ghost propagator $\vev{G_A}$ these equations are
shown in Fig.\ \ref{fig:dses}. It should be stressed that these Hamiltonian DSEs are not
equations of motion in the usual sense, but rather relations between the Green functions
and the so far undetermined variational kernels.


\section{Variational approach}

The Yang--Mills Hamilton operator in Coulomb gauge reads \citep{ChrLee80}
\begin{multline}\label{mad5}
H_\textsc{ym} =
\int \left[ - \frac{1}{2} \: J^{-1}_A \, \frac{\delta}{\delta A} \: J_A \frac{\delta}{\delta A} + \frac{1}{2} \: B^2 \right] \\
- \frac{g^2}{2} \int J^{-1}_A 
\biggl(\hat{A} \frac{\delta}{\delta A}\biggr) \, J_A \, F_A \, \biggl(\hat{A} \frac{\delta}{\delta A}\biggr),
\end{multline}
where $B$ is the non-abelian magnetic field, $\Hat{A}$ is the gauge field in the adjoint
representation of the colour group, and $F_A = G_A (-\partial^2) G_A $ is the Coulomb
interaction kernel. The vacuum energy is evaluated as VEV of the Hamilton operator \Eqref{mad5} with
the vacuum state defined by Eqs.\ \eqref{mad2} and \eqref{mad4}. By using the DSEs
stemming from the identity \Eqref{mad3}, the energy density can be written as a functional
of the variational kernels,
\be\label{mad5a}
\vev{H_\textsc{ym}} = E[\omega,\gamma_3] .
\ee
By using a skeleton expansion, the vacuum energy can be written at the desired order of
loops. Confining ourselves to two loops, the variation of the vacuum energy \Eqref{mad5a}
with respect to the kernel $\gamma_3$ fixes the latter to
\begin{multline}\label{mad6}
\gamma^{abc}_{ijk}(\vp,\vq,\vk) = 2 \, \i \, g \, f^{abc} \\
\times \frac{ \delta_{ij} (p-q)_k + \delta_{jk} (q-k)_i +\delta_{ki} (k-p)_j }{\Omega(\vp) + \Omega(\vq) + \Omega(\vk)} .
\end{multline}
Equation \eqref{mad6} is reminiscent of the lowest-order perturbative result \citep{Cam+09},
with the perturbative gluon energy $\abs{\vp}$ replaced by the non-perturbative one
$\Omega(\vp)$.

\begin{figure}
\includegraphics[width=\linewidth]{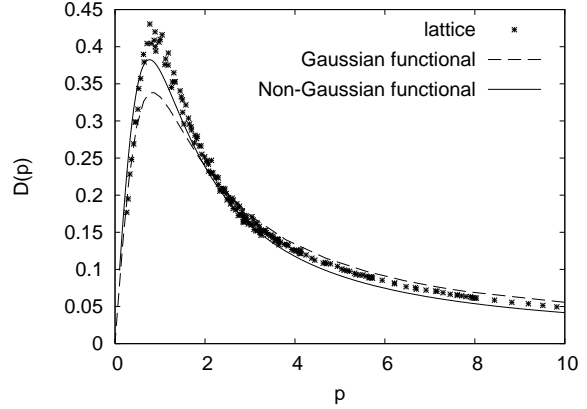}
\caption{\label{fig:prop}Gluon propagator obtained with a Gaussian (dashed line) and a
non-Gaussian functional (straight line), compared to the lattice data from Ref.\
\citep{BurQuaRei09}.}
\end{figure}

Combining the gluon DSE with the variational equation for the two-gluon kernel $\omega$,
one arrives at the gap equation for the gluon propagator
\be\label{mad7}
\Omega(\vp)^2 = \vp^2 + \chi(\vp)^2 + I_\mathrm{C}(\vp) - I_\mathrm{G}(\vp) .
\ee
The explicit expressions for the loop terms can be found in Ref.\ \citep{CamRei10}.
With a Gaussian wave functional, only the ghost loop $\chi(\vp)$ and the contribution
$I_\mathrm{C}(\vp)$ of the Coulomb kernel appear in the gap equation. The presence of
the gluon loop $I_\mathrm{G}(\vp)$ in the gap
equation \eqref{mad7} modifies the UV behaviour and allows us to extract, from the
non-renormalization of the ghost-gluon vertex, the correct first coefficient of the
$\beta$ function. The solution of the full set of coupled integral equations for the
ghost and gluon propagators and for the Coulomb form factor (see Ref.\ \citep{CamRei10})
is in progress. Here, in order to estimate the size of the gluon-loop contribution to
the gluon propagator, we use the gluon and ghost propagators obtained with a Gaussian
wave functional \citep{EppReiSch07} to calculate the gluon loop. The result is shown in
Fig.\ \ref{fig:prop}, together with lattice data from Ref.\ \citep{BurQuaRei09}.
The agreement between the continuum and the lattice results is improved in the
mid-momentum regime by the inclusion of the gluon loop, i.e.\ the three-gluon vertex,
as observed also in Landau gauge \citep{FisMaaPaw09}. The mismatch in the UV is a
consequence of the approximations involved, and should disappear when the full system
of coupled equations is solved.


\section{Three-gluon vertex}

\begin{figure}
\includegraphics[width=.6\linewidth]{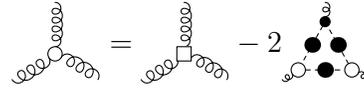}
\caption{\label{fig:3gvdse}Truncated DSE for the three-gluon vertex, under the assumption
of ghost dominance.}
\end{figure}

\begin{figure}
\includegraphics[width=\linewidth]{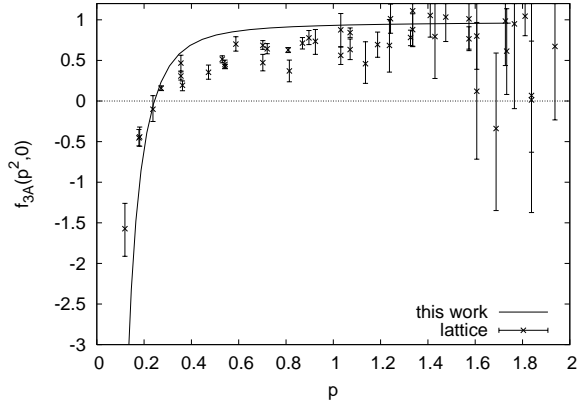}
\caption{\label{fig:3gv}Form factor $f_{3A}$ of the three-gluon vertex for orthogonal
momenta and comparison to lattice data for the $d=3$ Landau-gauge vertex \citep{CucMaaMen08}.
The momentum scale is arbitrary and has been adjusted to make the sign change occur at
the same point. The lattice data are shown by courtesy of A.~Maas.}
\end{figure}

The truncated DSE for the three-gluon vertex $\Gamma_3$ under the assumption of ghost
dominance is represented diagrammatically in Fig.\ \ref{fig:3gvdse}.
Possible tensor decompositions of the three-gluon vertex are given in Ref.\
\citep{BalChi80b}. Here, for sake of illustration, we confine ourselves to
the form factor corresponding to the tensor structure of the bare three-gluon vertex
\be\label{mad3gv1}
f_{3A} \coloneq
\frac{\Gamma_3 \cdot \Gamma_3^{(0)}}{\Gamma_3^{(0)} \cdot \Gamma_3^{(0)}},
\ee
where $\Gamma_3^{(0)}$ is the perturbative vertex, given by \Eqref{mad6} with $\Omega(\vp)$
replaced by $\abs{\vp}$. Furthermore, we consider a particular kinematic configuration,
where two momenta have the same magnitude
\be\label{mad3gv2}
\vp_1^2 = \vp_2^2 = p^2 , \quad \vp_1\cdot\vp_2 = c p^2 .
\ee
To evaluate the form factor $f_{3A}(p^2,c)$, we use the ghost and gluon propagators
obtained with a Gaussian wave functional \citep{EppReiSch07} as input. The IR analysis
of the equation for $f_{3A}(p^2,c)$ [\Eqref{mad3gv1}] performed in Ref.\
\citep{SchLedRei06} shows that this form factor should behave as a power law
in the IR, with an exponent three times the one of the ghost dressing function; this is
confirmed by our numerical solution \citep{CamRei10}. The result for the scalar form
factor $f_{3A}$ for orthogonal momenta, $f(p^2,0)$, is shown in Fig.\ \ref{fig:3gv},
together with lattice results for $d=3$ Landau gauge Yang--Mills theory.
Our result and the lattice data compare favourably in the low-momentum regime. In
particular, in both studies, the sign change of the form factor occurs roughly at the
same momentum where the gluon propagator has its maximum. (The scale in Fig.\ \ref{fig:3gv}
is arbitrary.)


\section{Conclusions}

We have presented a method to treat non-Gaussian wave functionals in the Hamiltonian
formulation of quantum field theory. By means of Dyson--Schwinger
techniques, the expectation value of
the Hamiltonian is expressed in terms of kernels occurring in the exponent of the
vacuum wave functional. These kernels are then determined by minimizing the vacuum
energy density. We have estimated the three-gluon vertex by using the propagators found
with a Gaussian wave functional as input. The result compares fairly well to the
available lattice data obtained in $d=3$ Landau gauge. The gap equation for the gluon
propagator contains the gluon loop, which was missed in previous variational approaches
with a Gaussian wave functional. The gluon loop gives a substantial contribution in the
mid-momentum regime while leaving the IR sector unchanged, and it also provides the
correct asymptotic UV behaviour of the gluon propagator in accord with perturbation
theory \citep{Cam+09}. The presently developed approach allows a systematic treatment of
correlators in the Hamiltonian formulation of a field theory, and opens up a wide range
of applications. In particular, it allows us to extend the variational approach from
pure Yang--Mills theory to full QCD.


\begin{theacknowledgments}
The authors are grateful to J.\ M.\ Pawlowski and P.\ Watson for useful discussions, and
to A.\ Maas for providing the lattice data from Ref.\ \citep{CucMaaMen08}. They also
thank the organizers for the interesting conference, and aknowledge financial support
by the DFG under contracts No.\ Re856/6-2,3 and by the Cusanuswerk.
\end{theacknowledgments}


\begin{thebibliography}{12}

\bibitem{SzcSwa01}
A.~P.~Szczepaniak, and E.~S.~Swanson, \emph{Phys. Rev.} \textbf{D65},
  025012 (2001), \texttt{hep-ph/0107078}.

\bibitem{FeuRei04}
C.~Feuchter, and H.~Reinhardt, \emph{Phys. Rev.} \textbf{D70},
  105021 (2004), \texttt{hep-th/0408236}.

\bibitem{EppReiSch07}
D.~Epple, H.~Reinhardt, and W.~Schleifenbaum, \emph{Phys. Rev.} \textbf{D75},
  045011 (2007), \texttt{hep-th/0612241}.

\bibitem{SchLedRei06}
W.~Schleifenbaum, M.~Leder, and H.~Reinhardt, \emph{Phys. Rev.} \textbf{D73},
  125019 (2006), \texttt{hep-th/0605115}.

\bibitem{YamSug10}
A.~Yamamoto, and H.~Suganuma, \emph{Phys. Rev.} \textbf{D81}, 014506 (2010),
  \texttt{0911.5391}.

\bibitem{CamRei10}
D.~R. Campagnari, and H.~Reinhardt, \emph{Phys. Rev.} \textbf{D}, in press
  (2010), \texttt{1009.4599}.

\bibitem{ChrLee80}
N.~H. Christ, and T.~D. Lee, \emph{Phys. Rev.} \textbf{D22}, 939--958 (1980).

\bibitem{Cam+09}
D.~Campagnari, A.~Weber, H.~Reinhardt, F.~Astorga, and W.~Schleifenbaum,
  \emph{Nucl. Phys.} \textbf{B842}, 501--528 (2011), \texttt{0910.4548}.

\bibitem{BurQuaRei09}
G.~Burgio, M.~Quandt, and H.~Reinhardt, \emph{Phys. Rev. Lett.} \textbf{102},
  032002 (2009), \texttt{0807.3291}.

\bibitem{FisMaaPaw09}
C.~S. Fischer, A.~Maas, and J.~M. Pawlowski, \emph{Annals Phys.} \textbf{324},
  2408--2437 (2009), \texttt{0810.1987}.

\bibitem{BalChi80b}
J.~S. Ball, and T.-W. Chiu, \emph{Phys. Rev.} \textbf{D22}, 2550 (1980).

\bibitem{CucMaaMen08}
A.~Cucchieri, A.~Maas, and T.~Mendes, \emph{Phys. Rev.} \textbf{D77}, 094510
  (2008), \texttt{0803.1798}.

\end{thebibliography}
\end{document}